\documentclass[journal=jctcce,manuscript=article]{achemso} %
\usepackage{float}              
\usepackage{supertabular}
\usepackage{longtable}
\usepackage{epsfig}
\usepackage{rotating}
\usepackage{exscale}            
\usepackage{graphicx}           
\usepackage{epstopdf}
\usepackage{amsfonts}           
\usepackage{amsmath}
\usepackage{amssymb}
\usepackage{ifthen}
\usepackage{xspace}
\usepackage{booktabs}
\usepackage[footnotesize]{subfigure}
\usepackage[dvips]{psfrag}
\usepackage{color,colortbl}
\usepackage{hyperref}
\providecommand{\url}[1]{\texttt{#1}}

\providecommand{\Capitalize}[1]{\uppercase{#1}}
\providecommand{\capitalize}[1]{\expandafter\Capitalize#1}

\usepackage[sc]{mathpazo}
\usepackage[T1]{fontenc}

\newcommand{\fref}[1]{Figure~\ref{#1}}
\newcommand{\tref}[1]{Table~\ref{#1}}

\newcommand{\rxn}[2]{{\bf #1} $\rightarrow$ {\bf #2}}

\newcommand{\eref}[1]{Equation~\eqref{#1}}

\makeatletter
\def\ps@paper{\def\@oddfoot{}\def\@oddhead{\hfil -- \thepage\ --\hfil}}
\title{Improving the Exploration of High Dimensional Free Energy Landscape by a Combination of Temperature Accelerated Sliced Sampling and Parallel Biasing}
\author{Abhinav Gupta}
\author{Shivani Verma} 
\author{Nisanth N. Nair}
\email{nnair@iitk.ac.in}
\affiliation{Department of Chemistry \\
Indian Institute of Technology Kanpur, 208016 Kanpur, India }

\keywords{Temperature Accelerate Sliced Sampling, Metadynamics, Parallel Bias Metadynamics, Umbrella Sampling, Weighted Histogram Analysis,  Free energy calculations, Alanine Pentapeptide}

\begin{document}
\thispagestyle{empty}

\maketitle

\clearpage

%\begin{document}
 
%\begin{tocentry}
%\includegraphics[height=3.5cm]{toc_mod.eps}
%\end{tocentry}

\begin{abstract}
Biased sampling methods such as the Temperature Accelerated Sliced Sampling (TASS), 
which can explore high dimensional collective variable (CV) space, is of great interest in free energy calculations.
Such methods can efficiently sample configurational space even when a large number of CVs for biasing are used while many conventional methods are limited to two or three CVs. 
In this paper, we propose a modification to the TASS method,
called Parallel Bias TASS or PBTASS, wherein a multidimensional parallel metadynamics bias is incorporated on a selected set of CVs. 
The corresponding time-dependent reweighting equations are derived, and the method is benchmarked.
In particular, we compare the accuracy and efficiency of PBTASS  with various methods viz. standard TASS, Temperature Accelerated Molecular Dynamics/driven-Adiabatic Free Energy Dynamics, and Parallel Bias Metadynamics. 
We demonstrate the capability of the PBTASS method by reconstructing the eight-dimensional free energy surface of alanine pentapeptide {\em in vacuo} from a 25 ns long trajectory.
Free energy barriers and free energies of high energy saddle points on the
high dimensional free energy landscape of this system are reported.
\end{abstract}

\section{Introduction}
%
% I think we should use \section*{}
%
Computational modeling of transitions from one metastable state to another on a free energy basin 
is of great interest in molecular simulations.
Such simulations are crucial in predicting the mechanism, kinetics, and thermodynamics of chemical reactions and  physical transformations.
%
%These tools are found to be useful in designing products with designed properties.
%
% What do you mean by designed properties?
%
%Molecular dynamics (MD) is one of the computational tools, which is widely used now a days to study physical processes at the molecular levels. Force fields are use to compute forces on the atoms felt by other surrounding atoms. 
% The time scale of the is limited in its use due to the limited time scale that makes it impossible to see any barrier crossing event termed as "rare event" \cite{Peters:Book,Tuckerman:Book,Vanden:2009:jcc,vanGunsteren:JCC:Rev:2010,Ciccotti:12,Pratyush:2016,shalini:review:2019}.
%
%
Often, the time-scale at which transitions occur among the metastable states is orders of magnitude larger than the period of bond vibrations in molecules. 
Thus such transitions are classified as rare-events in molecular simulations. 
Several enhanced sampling methods have been developed to accelerate such rare-events in molecular dynamics (MD)
simulations~\cite{Peters:Book,Tuckerman:Book,Vanden:2009:jcc,vanGunsteren:JCC:Rev:2010,Ciccotti:12,Pratyush:2016,shalini:review:2019,vashisth:review:2019}.

A typical approach to monitor the transitions among the metastable states is 
by looking at the progress along specific order parameters. 
The free energies 
computed along the order parameters can be used to calculate rates of the physio-chemical processes of interest.\cite{vashisth:review:2019}
In this spirit, collective variables (CVs) based enhanced sampling methods are proposed.\cite{Pratyush:2016}
In these methods, CVs, $\mathbf s(\mathbf R)$, 
which are functions of atomic coordinates $\mathbf{R}$, are considered and the quantity of interest is the Helmholtz free energy as a function of $\mathbf s$, given by,
\begin{eqnarray} 
   F (\mathbf s) &=& - k_{\rm B}T \ln P(\mathbf s)  \, . \label{f:p:1} \nonumber
\end{eqnarray} 
Here, $k_B$ is the Boltzmann constant, $T$ is the temperature of the system, and $\mathbf s \equiv (s_1,\cdots,s_n)$ is a vector in the CV-space. The probability distribution, $P(\mathbf s)$, is defined as 
\begin{eqnarray}
    P(\mathbf{s}^\prime) = \left \langle \prod_{i}^n \delta (s_{i}(\mathbf{R}) - s_{i}^\prime) \right \rangle 
    \nonumber
\end{eqnarray}
and can be calculated using a normalized histogram of $\mathbf s$ obtained from 
a canonical ensemble MD trajectory.
Enhanced sampling of CVs can be achieved by adding external bias potentials.
In umbrella sampling~\cite{us:orig,Kastner:11} 
(US), a harmonic bias  
\begin{eqnarray}
\label{eqn:wham_1}
    W^{\rm b}_h(s) = \frac{\kappa_h}{2} \left (s(\mathbf{R}) -\zeta_h \right )^2 ,   ~~~~ h = 1,\cdots,M,
\end{eqnarray}
is added to the system, where $\kappa_h$ is the force constant parameter and $\zeta_h$ is the 
mean position of the applied harmonic bias. 
Once the biased distributions centered at different regions in CV-space, $P^{\rm b}_h(s)$, $h=1,\cdots,M$, are obtained, they are reweighted and combined using the Weighted Histogram Analysis Method (WHAM).\cite{wham:1,wham:2}
In WHAM, the following equations are solved in a self-consistent (iterative) manner,
\begin{eqnarray}
\label{eqn:wham_2}
 P(s)= \frac {\sum_{h=1}^{M} n_h  P_h^{\rm b}(s)}  {\sum_{h=1}^{M} n_h \exp[\beta f_h] \exp[-  \beta  W_h(s)]}
\end{eqnarray}
where $f_h$ is computed at every step using $P(s)$ computed  
from the previous iteration as,
\[ \exp (- \beta f_h) = \int ds \exp \left [ -  \beta  W_h(s) \right ] 
 P(s) \enspace. \]
Here, $n_h$ is the number of frames in the $h^{\rm{th}}$ umbrella window.
 %
% The importance of umbrella sampling is that the sampling along any CV is done in a controlled fashion, but its sampling efficiency is limited due to its exponential dependency on the number of CVs, which limits the use of umbrella sampling to one dimension only.
 %
Often, the bias $W_h$ is one-dimensional, and seldom two-dimensional.
The advantage of the US method is that a controlled sampling is achievable
by the nature of the 
%, thanks to the 
restraining bias potential.
However, while dealing with large systems with several soft-modes, free energy convergence can be very slow,
%using the US.
%
%The slow convergence 
which can be attributed to the inadequate sampling of orthogonal coordinates.~\cite{shalini:review:2019,vashisth:review:2019}
%using a low-dimensional US bias
%

In metadynamics (MTD) \cite{Laio:PNAS:02,Iannuzzi:03} a time-dependent bias, $ V^{\rm b}(\mathbf{s},t)$, is added to the
potential,
\begin{eqnarray}
\label{eqn:MTD}
 V^{\rm b}(\mathbf{s},t) = \sum_{\tau < t }  w(\tau) \exp \left [ -\frac{( \mathbf s(\mathbf{R},t) - \mathbf s_\tau )^2}{2 (\delta s)^2} \right ]
 \end{eqnarray}
where $\tau$ runs over all the time-steps for  which the metadynamics bias was updated by augmenting a Gaussian centered at $\mathbf s_\tau \equiv \mathbf s(\mathbf R; \tau)$. 
Here, $\delta s$ and $w(\tau)$ are width and height parameters, respectively. 
In the conventional MTD approach, $w(\tau)$ is taken as a constant, $w(\tau)=w(0)$, while in the 
well-tempered version of MTD~\cite{mtd:well:08}, $w(\tau)$ is calculated as 
 \begin{eqnarray}
 \label{eqn:wtmtd}
     w(\tau) = w(0) \exp\left [ -\frac{V^{\rm b}(\mathbf{s},t)}{k_B \Delta T} \right ] \enspace . \enspace
 \end{eqnarray}
Here $w(0)$ is the initial height of the Gaussian function, and $V^{\rm b}(\mathbf s,t)$ is the bias 
%felt on the CVs 
at time $t$. 
Here $\Delta T$ is a parameter which controls the bias added at any time $\tau$ depending on the overall bias at $\mathbf s(\tau)$. 
The self-guiding nature of MTD makes it very efficient in studying complex chemical reactions and transformations. 
The method is widely used and excellent reviews are available.~\cite{Laio06:Book,mtd-rev06,MTD:Review:Gervasio,mtd:rev:11,Luigi:WIREs:MTDRev:2012,Bussi:Entropy,Bussi:Book:MTDRev:2015,Pietrucci:RP:2017,Allison:2020,Bussi2020}
However, the efficiency of the method 
decreases with the dimensionality of the CV-space. %
Thus the method is largely employed for problems where the dimensionality of the CV-space is 
not beyond three. 
Like in the case of the US method, slow convergence of free energy can be encountered when there is inadequate acceleration of (hidden) orthogonal coordinates.\cite{Pietrucci:RP:2017,shalini:review:2019}
%
%
%The sampling time for MTD also depends on the dimensionality of CV. It increases exponentially with the dimensionality of CV. As discussed for umbrella sampling, convergence of free energy is an issue with MTD because of the other orthogonal CVs \cite{shalini:review:2019}.

In order to overcome the problems due to insufficient sampling of transverse coordinates, the enhanced sampling methods must have the ability to accommodate a large number of CVs.
Most importantly, the performance of such methods should not deplete in an exponential manner on increasing the  dimensionality of the CV-space. 
Further, the method should permit the system to exhaustively transverse through relevant regions of a high dimensional CV-space and quickly provide a reliable estimate of the underlying free energy.
%
%A variant of the MTD method, namely 
The bias-exchange MTD (BEMTD)\cite{be:mtd:1, be:mtd:2} approach was put forward to achieve this within the MTD framework.
Here a certain number of replicas of the system are first created, and within each replica, a different set of CVs are enhanced sampled by low dimensional MTD bias. 
Further, exchanges between replicas are attempted using the Metropolis-Hastings scheme.
The method is widely used to study complex biological systems.
For a review of the technique and its applications, readers are directed to Ref.\cite{shalini:review:2019}.
%see \fref{be_mtd}.
%
%Each replica samples one or two dimensional CV-space, therefore sampling a high-dimensional space becomes highly efficient.
%
%By the virtue of this, the efficiency of sampling doesn't deteriorate exponentially with the number of CVs
%This method is known to be computationally expensive due to the less exchange probability among replicas.
%
One of the major drawbacks of this method is that its performance decreases when the distributions of replicas are poorly overlapping, thereby diminishing the exchange probability.

In a similar spirit, Pfaendtner and Bonomi have proposed an alternative MTD approach called Parallel Bias MTD~\cite{Pfaendtner:2015} (PBMTD). 
This method uses a single replica, while many low dimensional biases are applied on a set of CVs to 
enable extensive sampling of a high dimensional CV-space.
The technique uses a time-dependent bias potential of the form,
 \begin{eqnarray}
 \label{eqn:PBMTD1}
V^{\rm pb}(s_1,\cdots,s_n,t) = - \frac{1}{\beta}  \ln \sum_{i}^n \exp \left [-\beta V_i^{\rm b}(s_i,t) \right ] \enspace , 
 \end{eqnarray}
where $\beta = \left (k_{\rm B}T \right )^{-1}$ and $V_i^{\rm b}(s_i,t)$ is given by \eref{eqn:MTD}.
Further, the height of the Gaussian bias
along a dimension is modified based on the feedback from other dimensions, as
%by including a probability factor into the bias equation as following
\begin{eqnarray}
\label{eqn:PBMTD2}
 w_i(\tau) = w_i(0) \exp\left [ -\frac{V_i ^{\rm b}({s_i},t)}{k_B \Delta T} \right ] P_i(s_i) 
 \end{eqnarray}
 where 
\[ P_i(s_i) = \frac {\exp \left [ {-\beta V_i^{\rm b} (s_i,t)  }\right ]  } { \sum_{j}^n \exp{\left [  -\beta V_j^{\rm b}(s_j,t)\right ] }} \enspace , \mbox{and} \enspace \enspace i=1,\cdots,n \enspace . \]
Here $ P_i(s_i)$ is a feedback function to control the Gaussian height based on the bias deposited along other CVs.
The free energy along a CV $s_i$ can then be calculated as,
\begin{eqnarray}
\label{fes:pbmtd}
F_i(s_i) = -\gamma V^{\rm b}_i(s_i, t\rightarrow \infty )
\end{eqnarray}
where $\gamma = (T + \Delta T)/\Delta T$. 
This method is a significant improvement over the conventional MTD approach. 
However, very low values of $P_i$  with increasing dimensionality can lead to insignificant filling rate.
This problem can be addressed to some extent by selectively grouping the CVs.\cite{Pfaendtner:JCTC:2018b} 
The method is applied to 
a wide spectrum of complex chemical and biological problems; See Refs~\cite{Ferguson:PBMTD:2020,Pfaedtner:PBMTD:2020,Ricagno:PBMTD:2020,Shea:PBMTD:2020,Peter:PBMTD:2020} and references in Ref.~\cite{shalini:review:2019}.
%
%probability factor which makes the process of adding bias very slow and thus sampling of free energy surface is also limited. 
%However, the number of CVs can be increased but dimensionality of the added bias should be kept low. 
%
%Though it seems that any number of CVs can now be biased using parallel bias MTD but it should kept in mind that the probability factor plays a major role in biasing. The more the number of CVs the slower is the sampling. It may solve the problem of biasing more number of CVs at a time but it is computationally inefficient due to its sampling time. \\
%

Several other attempts to improve the sampling using a MTD-like bias have been also put 
forward lately.\cite{Parrinello:JPCL:2020,Parrinello:JCTC:2019,Sugita:2019,Valsson2020,Bussi2020} 

Temperature accelerated molecular dynamics/driven-adiabatic free energy dynamics\cite{Tuckerman:Book,TAMD:1,Tuckerman:2008} is a powerful approach 
for exploring high dimensional free energy landscapes.\cite{Rosso:JCP:2002,AFED2}
Hereafter we will denote this method as TAMD. 
Accelerated diffusion of the system in the CV-space is achieved in this method by increasing the temperature of the fictitious degrees of freedom that are restrained to the CVs.
%The force field should be checked for its applicability at high temperature beforehand. 
TAMD employs the extended Lagrangian,
\begin{eqnarray}        
\label{eqn:tamd:lag}
\mathcal L_{\rm TAMD}(\mathbf R, \dot{\mathbf R}, \mathbf z,\dot{\mathbf z}) = \mathcal L_{0}(\mathbf R, \dot{\mathbf R}) + \sum_{i=1}^{n} \frac {1}{2} \mu_{i} \dot{\mathbf z}_{i}^2 - \nonumber \sum_{i=1}^{n} \frac{k_{i}}{2} ( s_{i}(\mathbf R) - z_{i})^2  \enspace
\end{eqnarray}  
where $\mathcal L_{0}$ is the Lagrangian of the physical system,  $\{z_i\}$ is the set of $n$ fictitious variables corresponding to $n$ CVs $\{s_i(\mathbf R) \}$, $\mu_i$ is the mass of the auxiliary variable $z_i$, and 
$k_{i}$ is the harmonic spring constant 
in the potential that restrains the motion of $z_i$ and $s_i$. 
Auxiliary variables are set to a high temperature $\tilde T$, which is much higher than
the temperature $T$ of the physical system.
The parameters $\{\mu_{i}\}$ and $\{k_{i}\}$ 
are chosen in such a way that the adiabatic separation between the $\{s_i\}$ and $\{z_i\}$ degrees of freedom is maintained. 
%in order to ensure no heat flow between the two subsystems.
%
Separate thermostats are used to maintain the temperature of the physical system and the extended system. 
%
%This method is efficient and less expensive from other global tempering methods \cite{REM:1} where the whole system is heated in many replicas and their exchange is attempted. The free energy is obtained by simply calculating the probability distribution at high temperature $\tilde T$ and reweight
% reweight is correct word or not???
%to get the distribution at normal temperature $T$ \cite{Rosso:JCP:2002,Tuckerman:Book,TAMD:1} as follows
%
Free energy surface $F(\mathbf z)$ can be computed from the probability distribution of the auxiliary variables $\tilde P(\mathbf z)$ as,
\begin{eqnarray}
\label{eqn:tamd:fes}
 F(\mathbf z) = - k_{\rm B} \tilde T \ln \tilde P(\mathbf z) 
\end{eqnarray}
and is a good estimator for the underlying free energy surface $F(\mathbf s)$ 
along the physical coordinates at temperature $T$.\cite{Tuckerman:Book}
To further improve the efficiency of this method, a variant of the approach, called Unified Free Energy Dynamics (UFED), 
was proposed by Tuckerman and co-workers.\cite{Tuckerman:2012} 
For a review of TAMD and related methods, see Refs.~\cite{shalini:review:2019,vashisth:review:2019}.
%~\cite{}

To introduce a more controlled exploration of high dimensional free energy surfaces,
Temperature Accelerated Sliced Sampling (TASS) approach was introduced.\cite{Shalini:2017}
This method is built on the TAMD Lagrangian wherein a combination of US and MTD biases are applied on a selected 
set of CVs.
%in order to improve the sampling.
%
The Lagrangian used in TASS is,
\begin{eqnarray}
  \mathcal L_{\rm TASS}(\mathbf R,\dot{\mathbf R},\mathbf z,\dot{\mathbf z}) = 
  \mathcal L_{\rm TAMD}(\mathbf R,\dot{\mathbf R},\mathbf z,\dot{\mathbf z}) -  
  \frac{1}{2}\kappa_h \left ( z_1 - \zeta_h \right )^2  - V^{\rm b}(\overline{\mathbf z}, 
  t) \enspace , \enspace \enspace h=1,\cdots,M. 
\end{eqnarray}
Here US bias is applied along the auxiliary variable $z_1$, and such 
$M$ different biases centered at $\zeta_1,\cdots,\zeta_M$ are taken.
The restraining force constant of the US bias centered at $\zeta_h$ is $\kappa_h$. 
The metadynamics bias $V^{\rm b}(\overline{\mathbf z})$ acting on a subset of the 
auxiliary space $\overline{\mathbf z} \equiv
\left ( z_2,\cdots,z_m \right )$, with $m \leq n$ can be optionally considered as well.
The probability distributions of auxiliary variables obtained with different biases  
are then reweighted and combined to get the full high dimensional free energy landscape.\cite{Shalini:2017,shalini:review:2019}
%
%By using the framework which combines a one-dimensional US bias and TAMD/AFED type Lagrangian, 
%TASS gives flexibility in selecting transverse CVs to $s_1$ 
Other than the benefit of achieving a controlled sampling along $s_1$,
TASS provides flexibility in selecting different transverse CVs 
depending on the window $h$.
Further, a large 
number of orthogonal CVs can be chosen by virtue of the temperature acceleration of $\mathbf z$.
Different TASS windows can run in parallel, making the computations very efficient. 
%of different slices of high dimensional free energy landscape can be done in parallel 
%and an 
Each window samples a high dimensional slice of the free energy landscape,
thereby an exhaustive exploration is achieved through the divide-and-conquer strategy inherent to the TASS method.
It was observed that  TASS could obtain a quick convergence in free energy barriers.\cite{Shalini:2017,Awasthi:2018,VITHANI:2018,Sahoo:2018,soniya:acs:2019,anji:JICS:2019}
For a review of the method, see Refs.~\cite{vashisth:review:2019,shalini:review:2019}.

In the earlier applications of TASS, only 
a one-dimensional MTD bias was used.
Increasing the dimensionality of the MTD bias
decreases the performance of the method.
In this work, we propose a modified TASS approach, called Parallel Bias TASS (PBTASS), 
to  improve the efficiency of the method further.
This is accomplished by replacing the one-dimensional MTD bias used in the conventional TASS method by a 
high dimensional PBMTD bias.
A modified TASS reweighting scheme,
accounting the PBMTD bias, has been put forward.
The method is benchmarked for its accuracy and efficiency. 
Finally, PBTASS is used to explore the eight-dimensional free energy landscape of alanine pentapeptide ({\em in vacuo}).

\section{Theory}

%\subsection{Parallel Bias TASS Method}
We introduce the Parallel Bias TASS (or PBTASS) method that incorporates PBMTD bias
within TASS as follows:
\begin{eqnarray}
\label{PBTASS:eqn}
  \mathcal L_{{\rm PBTASS},h}\left ( \mathbf R, \dot{\mathbf R}, \mathbf z, \dot{\mathbf z} \right ) = 
  \mathcal L_{\rm TAMD}(\mathbf R,\dot{\mathbf R},\mathbf z,\dot{\mathbf z}) -  
  \frac{1}{2}\kappa_h \left ( z_1 - \zeta_h \right )^2  -  V^{\rm pb}(\overline {\mathbf z},t) \enspace , \enspace \enspace h=1,\cdots,M. 
\end{eqnarray}
The PBMTD bias $V^{\rm pb}(\overline{\mathbf z},t)$, as given by \eref{eqn:PBMTD1},  is applied 
on a subset of auxiliary variables $\overline{\mathbf z} \equiv (z_2,\cdots,z_m)$, with $m\leq n$.
All the variables ${\mathbf z} \equiv ( z_1,\cdots,z_n)$ are coupled to a massive thermostat
at temperature $\tilde T$, and $\tilde T >> T$. 
Then, $M$ independent simulations using the PBTASS Lagrangian (\eref{PBTASS:eqn}) are carried out after 
equilibrating the starting structure with the US bias (\eref{eqn:wham_1}).
%
%It is stressed here that the set of orthogonal CVs used for each  depending
%on the problem in hand.
%
{The PBTASS Lagrangian can be setup in a straightforward 
manner using the recent version of the PLUMED Interface,\cite{plumed:2009} and a sample input file is given in the Supporting Information.}

From the trajectories of these simulations, probability distributions of the CVs, $\tilde P_h(\mathbf z)$, for  $h=1,\cdots,M$ simulations, are computed by
binning.
Subsequently, these biased distributions are reweighted for the bias potential
$V^{\rm pb}(\overline{\mathbf z},t)$.
Several, but related, reweighting approaches\cite{Bonomi:09,Tuckerman:JCTC:2014,Tiwary:14,Salvalaglio:JCP:2019,Parrinello:JPCL:2020,Ceriotti:Reweight:2020} are available, while for our purpose, we derived a time-dependent reweighting scheme by taking a cue from the work of Tiwary and Parrinello.\cite{Tiwary:14}

If $\tilde F(\mathbf z)$ is the underlying multidimensional free energy
surface at temperature $\tilde T$, then the biased distribution obtained from a PBTASS simulation corresponding to the window $h$ can be written as,
\begin{eqnarray}
\tilde P_h^{\rm b}(\mathbf z,t) &=& \frac{
\exp\left \{ - \tilde \beta \tilde F(\mathbf z) \right \} 
\exp\left \{ - \tilde \beta V^{\rm pb}(\overline{\mathbf z},t) \right \} 
}
{
\int d \mathbf z \, \exp\left \{ - \tilde \beta \left [ 
\tilde F(\mathbf z) +  V^{\rm pb}(\overline{\mathbf z},t) \right ]
\right \}
} \nonumber \\[1ex]
&=& \frac{
        \exp\left \{ - \tilde \beta \tilde F(\mathbf z) \right \} 
        \exp\left \{ - \tilde \beta V^{\rm pb}(\overline{\mathbf z},t) \right \} 
       }
       { 
        \sum_{j=2}^m \int d \mathbf z \, 
        \exp\left \{ - \tilde \beta \tilde F(\mathbf z) \right \} 
        \exp\left \{ - \tilde \beta  V^{\rm b}_j(z_j,t ) \right \}
       } \nonumber \\ [1ex] 
&=& \frac{
        \exp\left \{ - \tilde \beta \tilde F(\mathbf z) \right \} 
        \exp\left \{ - \tilde \beta V^{\rm pb}(\overline{\mathbf z},t) \right \}
       }
       { 
        \sum_{j=2}^m \int d z_j \,  
         \exp\left \{ - \tilde \beta \tilde F_j(z_j) \right \}
        \exp\left \{ - \tilde \beta  V_j^{\rm b}(z_j,t ) \right \}
        }     \label{prob:bias1}  
\end{eqnarray}
where  $\tilde \beta = 1/(k_{\rm B} \tilde T)$, and 
$\tilde F_j(z_j)$ is the projection of $\tilde F(\mathbf z)$ along $z_j$. %
The US bias reweighting, which is not considered while deriving the above equation, will be integrated when using the WHAM at the final stage.
Here $j$ runs from 2 to $m$, because the first auxiliary variable, $z_1$ (corresponding to the CV $s_1$),
is biased by restraining potential, as shown in \eref{PBTASS:eqn}.
It is also emphasized here that $m \leq n$ because it is not necessary that all the other auxiliary coordinates
have to be biased by PBMTD.
In the above steps, we used the identity in \eref{eqn:PBMTD1}.
Unbiased probability distribution $\tilde P_h$ at temperature $\tilde T$ is given by,
\begin{eqnarray}
\tilde P_h(\mathbf z) = \frac{\exp\left \{ - \tilde \beta \tilde F(\mathbf z) \right \} }
{
\int d \mathbf z \, \exp\left \{ - \tilde \beta \tilde F(\mathbf z) \right \} 
} \enspace .   \label{pz:eqn}
\end{eqnarray}
Using \eref{pz:eqn}, we substitute for $\exp\left \{ - \tilde \beta \tilde F(\mathbf z) \right \} $ in \eref{prob:bias1} to yield,
\begin{eqnarray}
\tilde P_h^{\rm b}(\mathbf z,t) &=& \tilde P_h(\mathbf z) 
\frac{ \exp\left \{ - \tilde \beta V^{\rm pb}(\overline{\mathbf z},t) \right \} 
\int d \mathbf z \, \exp\left \{ - \tilde \beta \tilde F(\mathbf z) \right \}
}{
        \sum_{j=2}^m \int d z_j \,  
         \exp\left \{ - \tilde \beta \tilde F_j(z_j) \right \}
        \exp\left \{ - \tilde \beta  V_j^{\rm b}(z_j,t ) \right \}
        }   \nonumber  \\[1ex]
&=&  \tilde P_h(\mathbf z)
 \frac{ \exp\left \{ - \tilde \beta V^{\rm pb}(\overline{\mathbf z},t) \right \}
}{
        \sum_{j=2}^m  \left [ \int d z_j \,  
         \exp\left \{ - \tilde \beta \tilde F_j(z_j) \right \}
        \exp\left \{ - \tilde \beta  V_j^{\rm b}(z_j,t ) \right \} / Z_j \right ]
        }  \nonumber
%        \label{prob:bias2}  
\end{eqnarray}
where $Z_j=\int d z_j \, \exp\left \{ - \tilde \beta \tilde F_j(z_j) \right \}$.
On rearranging the above equation, we get the expression for the unbiased
probability distribution as,
\begin{eqnarray}
\tilde P_h(\mathbf z) &=& \tilde P_h^{\rm b}(\mathbf z,t) 
\frac{
        \sum_{j=2}^m  \left [ \int d z_j \,  
        \exp\left \{ - \tilde \beta \tilde F_j(z_j) \right \}
        \exp\left \{ - \tilde \beta  V_j^{\rm b}(z_j,t ) \right \} / Z_j \right ]
        } 
        {\exp\left \{ - \tilde \beta V^{\rm pb}(\overline{\mathbf z},t) \right \}
        } \nonumber \\ [1ex]
        &=& \tilde P_h^{\rm b}(\mathbf z,t) \exp\left \{ \tilde \beta \left [  V^{\rm pb}(\overline{\mathbf z},t)  + c(t) \right ] \right \}
\end{eqnarray}
with 
\begin{eqnarray}
\exp\left \{ \tilde \beta c(t) \right \} &=& \sum_{j=2}^m  \left [ 
\exp\left \{ - \tilde \beta \tilde F_j(z_j) \right \}
\exp\left \{ - \tilde \beta  V_j^{\rm b}(z_j,t )  \right \} / Z_j \right ] \nonumber  \\ [1ex] 
&\approx& 
\sum_{j=2}^m \frac{
\int d z_j \,  \exp\left \{  \tilde \beta \gamma V_j^{\rm b}(z_j,t) \right \} \exp\left \{ - \tilde \beta  V_j^{\rm b}(z_j,t ) \right \}
}{\int d z_j \, \exp\left \{ \tilde \beta \gamma V_j^{\rm b}(z_j,t) \right \} }
\nonumber
\end{eqnarray}
where we used \eref{fes:pbmtd} and  $\gamma = (\tilde T + \Delta T)/\Delta T$. 
The last equation becomes exact in the limit $t \rightarrow \infty$.
Thus, we obtain the relation,
\begin{eqnarray}
c(t) = \tilde{\beta}^{-1} \ln \left [ 
\sum_{j=2}^m  \frac{\int d z_j \, 
\exp \left \{ {\tilde \beta \left ( \gamma - 1 \right ) V^{\rm b}(z_j,t) } \right \}  }{ \int d z_j \exp \left \{ \tilde \beta \gamma V^{\rm b}(z_j, t) \right \} }
\right ]   \enspace . \nonumber
\end{eqnarray}

%Following the Tiwary-Parrinello reweighting scheme\cite{Tiwary:14}, it is straightforward to show that this can be done by,
%\nncomment{It is better to show the derivation}
%

For each window $h$, we can thus obtain the  
distribution 
\begin{eqnarray}  
\label{eqn:reweight:tiwary} 
\tilde P_h(\mathbf z^\prime) = 
\left < \prod_i^n \delta(z_i - z_i^\prime) \, \exp \left \{ \tilde \beta \left [ V^{\rm pb}( \overline{\mathbf z},t) + c(t) \right ] \right \}  \right >  \enspace . \nonumber \end{eqnarray}
%
%and 
%\begin{eqnarray}
%c(t) = \tilde{\beta}^{-1} \ln \left [ 
%\sum_{i=2}^m  \frac{\int d z_i \, 
%\exp \left \{ {\tilde \beta \left ( \gamma - 1 \right ) V^{\rm b}(z_i,t) } \right \}  }{ \int d z_i \exp \left \{ \tilde \beta \gamma V^{\rm b}(z_i, t) \right \} }
%\right ]   \enspace ,
%\end{eqnarray}
%where $\tilde \beta = 1/(k_{\rm B} \tilde T)$, and $\gamma = (\tilde T + \Delta T)/\Delta T$.
%
In practice, the above can be implemented by a time-dependent
binning of the time series data of the auxiliary variables as,
%this is done by a time-depeg ndent reweighting employing 
%pb}(\overline{\mathbf z},t) + c(t) \right]
%\right \}$. 
%\nncomment{better to write the time dependent reweighting expression %here}
%
\begin{eqnarray}
\tilde P_h(\mathbf z_i^\prime) =  \frac { \int_{\rm t_{min}}^{\rm t_{max}} d \tau  \exp \left \{ \tilde{\beta} \left [ V^{\rm pb} (\overline{\mathbf z},t) + c(t) \right] \right \}  \prod_{i=1}^{n} \delta(\mathbf{z}_i - \mathbf z_i^\prime) } 
{  \int_{\rm t_{min}}^{\rm t_{max} } d \tau  \exp \left \{ \tilde{\beta} \left [ V^{\rm pb}(\overline{\mathbf z},t) + c(t) \right] \right \} }
\enspace .  \label{eqn:pbtass:reweight}
\end{eqnarray}
In our calculations, $t_{\rm min}$ was set as the first time-step of PBTASS and
$t_{\rm max}$ was varied till a satisfactory convergence in free energy estimates was observed.
Although varying $t_{\rm min}$ didn't make any difference in our calculations,
it may become important to choose a suitable value when PBTASS simulation is not started from a good initial structure and when the initial bias growth rate is very high.

Subsequently, $M$ distributions $ \{ \tilde P_h\}$, as obtained using \eref{eqn:pbtass:reweight}, are combined to get $\tilde P(\mathbf z)$ using WHAM, exercising \eref{eqn:wham_2}, and
the free energy surface $F(\mathbf z)$ for temperature $T$ is  computed using \eref{eqn:tamd:fes}.
For bias reweighting and performing WHAM, we have developed  our own programs.

It is better to perform WHAM on low
dimensional distributions.
For this purpose, the probability distribution of each slice may be projected to a set of relevant low dimensions before carrying out WHAM.
A more general mean-force-based approach\cite{Chipot:2015,E:JCP:2014,Tuckerman:2015} can be formulated to combine the free energy slices in this case, thereby evading WHAM.
This will be communicated in a forthcoming publication.

%The application of the PBTASS method is done to sample the configuration space of two real small peptide systems in vacuum, alanine tripeptide and alanine pentapeptide, to validate the method. All ($ \phi, \psi$) torsional angles are used as CVs for both the systems. The results are compared with the conventional parallel bias metadynamics (PB-MTD) \cite{Pfaendtner:2015}, temperature accelerated molecular dynamics (TAMD) and the original TASS method.   
% 
%
\section{Results and Discussion}

\subsection{Alanine Tripeptide {\em In Vacuo}}
\begin{figure}[htbp]
\centering
    \includegraphics[scale=0.15]{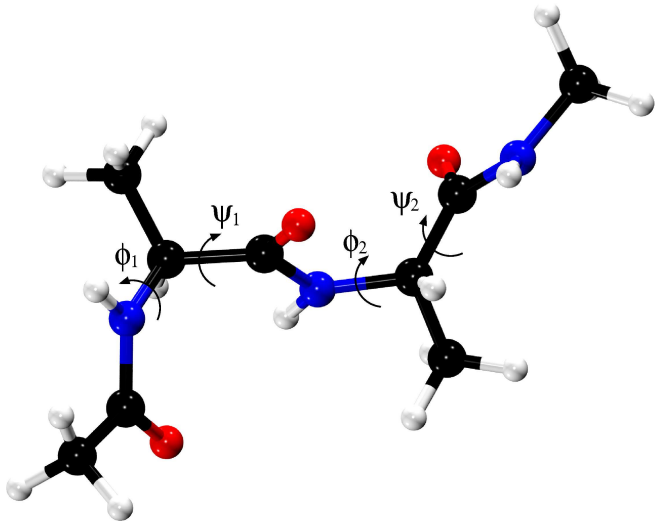}
     \caption{\label{tri-ala}
     Alanine tripeptide molecule is shown where the four Ramachandran angles 
     ($\phi_1,\psi_1,\phi_2,\psi_2$) chosen as CVs are labelled. Atom color codes: Black (C); Blue (N); Red (O); White (H).
     }
     \end{figure}

At first, we investigated the free energetics of 
alanine tripeptide {\em in vacuo} to benchmark the PBTASS method. 
MD calculations were performed using AMBER14\cite{ff14SB} %TODO Write the version number instead of XX
interfaced with PLUMED-2.2.3.\cite{plumed2.2.3}
The ff14SB force field~\cite{ff14SB} was taken to describe interatomic interactions. 
We chose four Ramachandran angles ($\phi_1,\psi_1,\phi_2,\psi_2$) as CVs and the free energy surface $F(\phi_1,\psi_1,\phi_2,\psi_2)$ was computed using PBTASS, TASS, TAMD, and PBMTD methods; see  ~\fref{tri-ala} for  the definition of CVs.

In PBTASS simulations, we chose $\phi_1$ for applying the umbrella bias.
A three-dimensional PBMTD bias, $V^{\rm pb}(\psi_1,\phi_2,\psi_2)$, was applied along the CVs 
$\psi_1$, $\phi_2$, and $\psi_2$.
As in \eref{eqn:PBMTD1}, the three-dimensional bias was constructed based on the three one-dimensional MTD biases $V^{\rm b}(\psi_1)$, $V^{\rm b}(\phi_2)$, and $V^{\rm b}(\psi_2)$.
The choice of the type of the bias applied along a CV was arbitrary for this problem.
%
%
%The simulations setup was designed with these four CVs: umbrella sampling was performed along $\phi_1$ and metadynamics parallel biases were applied along ($\psi_1, \phi_2, \psi_2$). 
%Auxiliary variables corresponding to the four CVs were thermostatted to 3000~K. 
%
An overdamped Langevin thermostat with a friction coefficient of
0.1~fs$^{-1}$ was used to maintain the auxiliary variable temperature at 3000 K.
%
%The performance of force field at this temperature was also checked by the deviation of their fluctuations of bonds, angles and dihedrals. 
The PBMTD bias potential was updated every 500~fs, and the bias parameters $w_1(0) = w_2(0) = w_3(0) =0.6$~kcal~mol$^{-1}$ and $\delta s =$0.05 radians were taken. 
The parameter {$\Delta T$ was set to $21000$~K } in order to achieve a reasonable bias growth rate. 
The umbrella potential was applied from $-\pi$ to $\pi$ at an interval of 0.2 radians with $\kappa_h = 1.2\times10^2 $ kcal~ mol$^{-1}$rad$^{-2}$,  $k_i=1.2\times 10^3$~kcal~mol$^{-1}$rad$^{-2}$, and  $\mu_i=50$~amu {\AA}$^2$ rad$^{-2}$,
for $h=1,\cdots,33$ and $i=1,\cdots,4$.
%
%Here, {\tt \color{Red} $M=33$ }
%and {\tt \color{red} $n=4$}.
Langevin thermostat with a friction coefficient of 0.1~fs$^{-1}$ was used for maintaining the temperature of physical system at 300~K. 
Before starting the PBTASS simulation for a specific window, we carried out equilibration for 100 ps.
Starting structures for the equilibration runs were taken as the
global minimum structure of alanine tripeptide for all the windows.

\begin{figure*}
\centering
    \includegraphics[width=\textwidth]{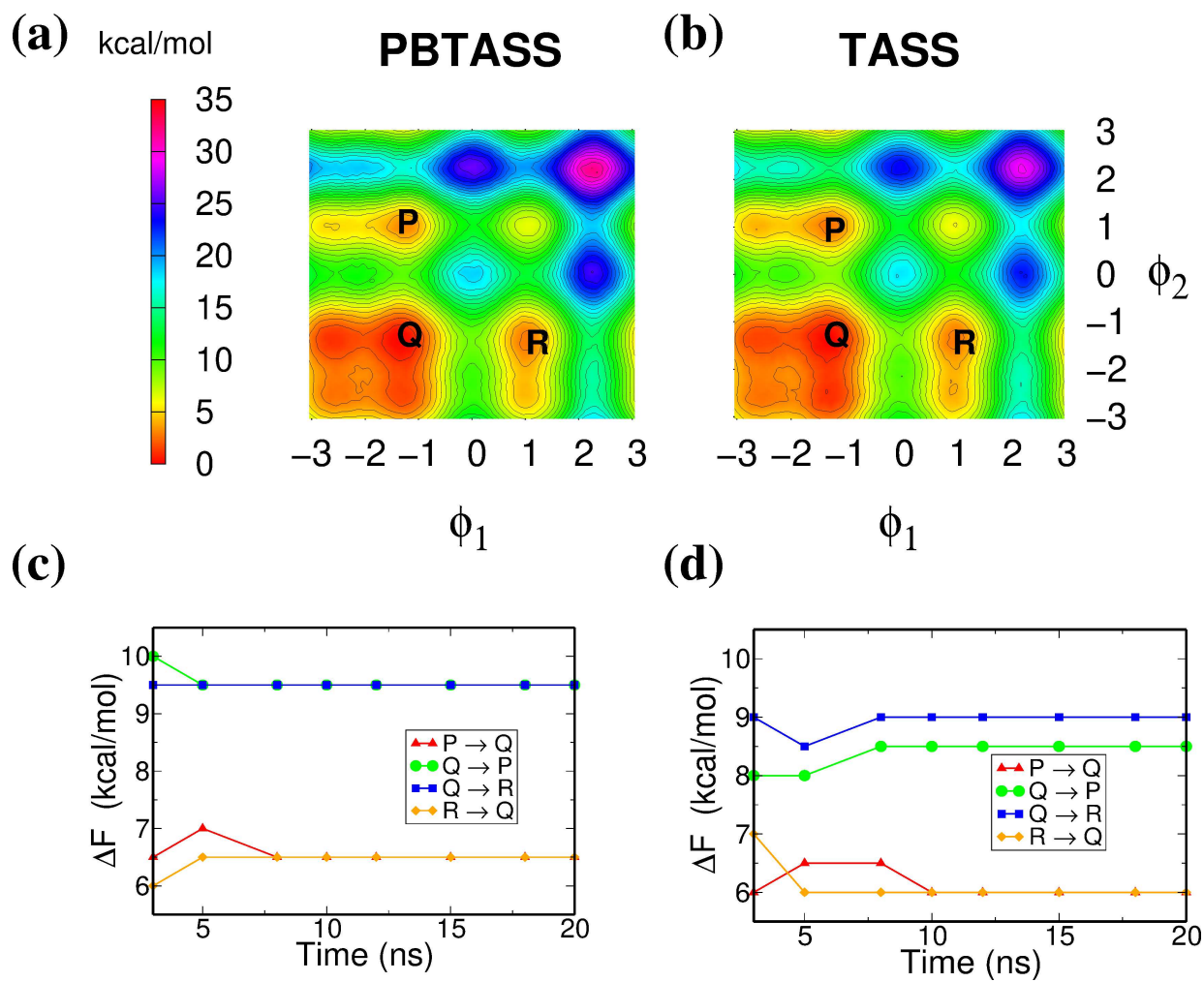}
    \caption{The free energy surface $F(\phi_1,\phi_2)$ of alanine tripeptide {\em in vacuo} computed from (a) PBTASS and (b) TASS simulations after 20~ns 
     per window. Contours are drawn for every 1 kcal~mol$^{-1}$. Convergence of various free energy barriers on these surfaces 
    as a function of simulation time per window
    is shown in the lower panels (c) and (d).
    }
     \label{fig:tri-ala1.0}
\end{figure*}

\begin{figure*} [h]
\centering
    \includegraphics[width=\textwidth]{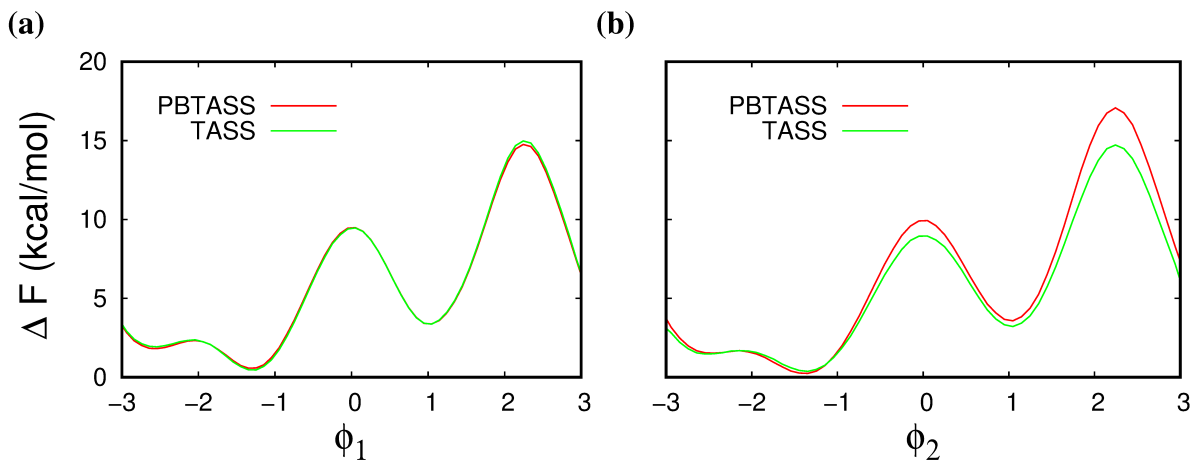}
    \caption{(a) $F(\phi_1)$, and (b) $F(\phi_2)$ for alanine tripeptide {\em in vacuo} computed from PBTASS (red) and TASS (green) simulations after 20~ns 
     per window. Here angles are in radians. 
    }
     \label{fig:tri-ala1.1}
\end{figure*}

\begin{figure*}
\centering
    \includegraphics[width=\textwidth]{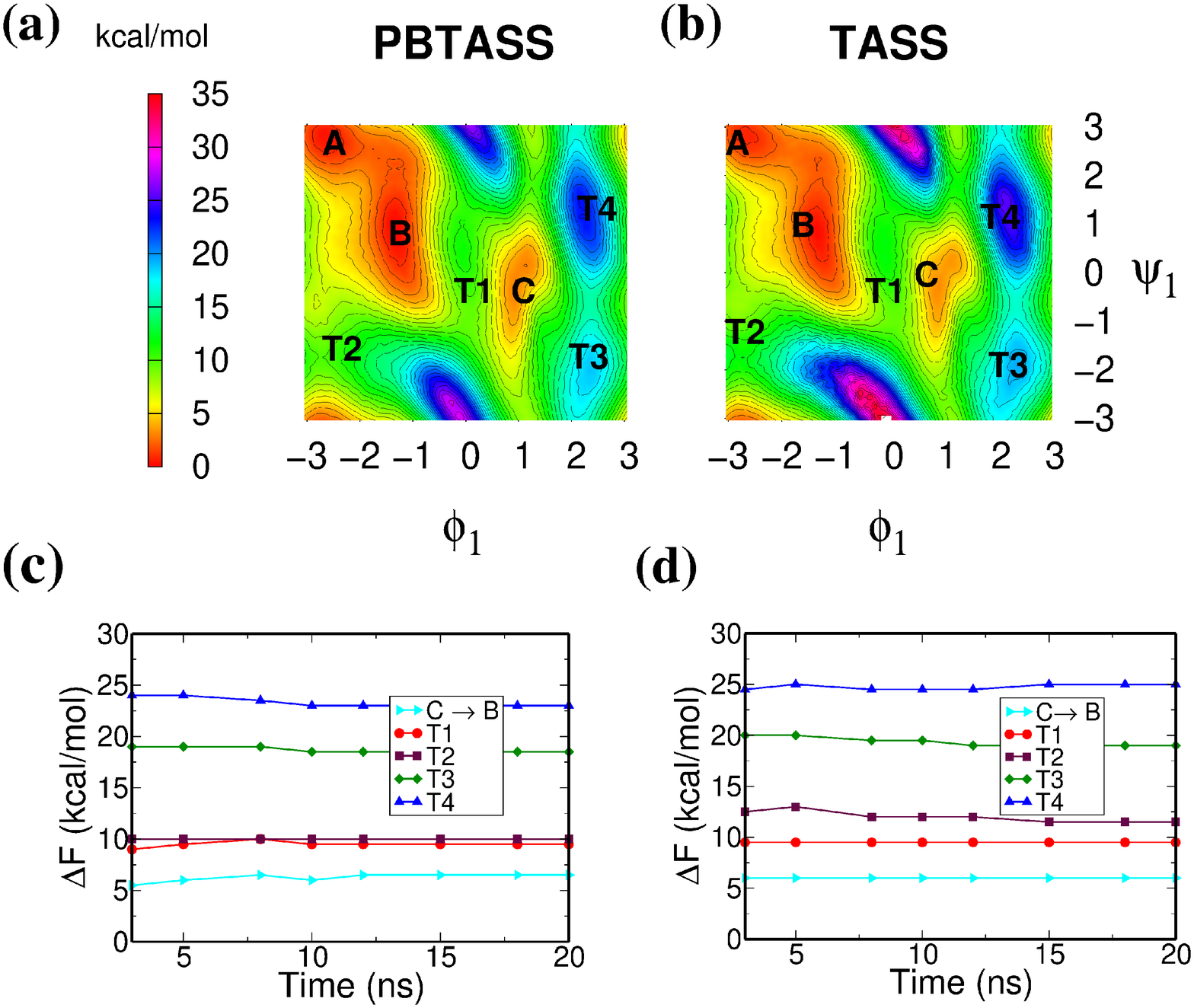}
    \caption{The free energy surface $F(\phi_1,\psi_1)$ of alanine tripeptide {\em in vacuo} computed from (a) PBTASS and (b) TASS simulations after 20~ns 
     per window.
    Contours are drawn for every 1 kcal~mol$^{-1}$.
    Convergence of free energy barrier for \rxn{B}{C}, and free energies of the 
    saddle points {\bf T1}, {\bf T2}, {\bf T3}, and {\bf T4} with respect to the free energy of {\bf B} 
    as a function of simulation time per window
    in PBTASS (c), and TASS (d) runs are also shown in the lower panels.  Here
    angles are in radians.
     \label{fig:tri-ala3}}
\end{figure*}

%MOVE THIS FIGURE TO SI - THIS WILL BE SI~Figure~1.
%\begin{figure*}
%\centering
%    \includegraphics[width=\textwidth]{ala_tri_phi2_psi2_com.eps}
%    \caption{The free energy surface $F(\phi_2,\psi_2)$ of alanine tripeptide {\em in vacuo} computed from (a) PBTASS and (b) TASS simulations after 20~ns runs per umbrella window. Convergence of free energy barrier for \rxn{B'}{C'} and  free energies of saddle points {\bf T1'}, {\bf T2'}, {\bf T3'}, and {\bf T4'} with
%    respect to the free energy of {\bf A'} 
%    as a function of simulation time per umbrella window in PBTASS (c), and TASS (d) runs.  Here angles are in radians.}
%     \label{fig:tri-ala4}
%\end{figure*}

%To validate and benchmark the PBTASS scheme, we performed TASS, TAMD, and PBMTD simulations
%for this system. 
%
For TASS, TAMD, and PBMTD runs, we used identical set up as that of the PBTASS simulation.
In TASS, the MTD bias (\eref{eqn:MTD}) was applied only along $\phi_2$. 
For the benefit of comparison, the parameters used in these simulations were kept the same as that used in PBTASS runs.
%
%Similarly, in TAMD and PBMTD simulations, the details of system setup and parameters used were the same as that of the PBTASS simulation.
%
In PBMTD reference simulations, bias potential was applied only along
$\phi_1$ and $\phi_2$.

Four-dimensional free energy surface $F(\phi_1,\psi_1,\phi_2,\psi_2)$ was computed from the PBTASS trajectory using the method described in Section~2, and the high dimensional surface was projected on the $(\phi_1, \phi_2)$ space for analysis; See \fref{fig:tri-ala1.0}.
The three main metastable states on the $F(\phi_1,\phi_2)$ free energy surface are labelled as {\bf P}, {\bf Q}, and {\bf R}.
The converged free energy barriers {\bf Q}$\rightarrow${\bf R} and {\bf Q}$\rightarrow${\bf P} are { 9.5~kcal~mol$^{-1}$, while the reverse barriers, i.e. {\bf P}$\rightarrow${\bf Q} and {\bf R}$\rightarrow${\bf Q}, are 6.5~kcal~mol$^{-1}$}.
Of great importance, the free energy barriers are converged within $\sim$8~ns (per window).
It is also gratifying to see that the diagonal reflection symmetry\cite{Shalini:2017} of the $F(\phi_1,\phi_2)$ surface is retained.

On the other hand, the free energy barriers computed using TASS are converged to values within $\sim$1~kcal~mol$^{-1}$ of  that computed from PBTASS; See also \tref{table:ala3}.
Ideally, the {\bf Q}$\rightarrow${\bf R} and {\bf Q}$\rightarrow${\bf P} barriers have to be the same (due to the symmetry of the surface), however, a small difference ($\sim$0.5~kcal~mol$^{-1}$) is noticed. 
This could be due to poor convergence of free energy along $\phi_2$, 
where the MTD bias potential with 
a high value of $\Delta T$
was applied in TASS.
After 10~ns, reverse barriers were found to 
be the same and equal to 6~kcal~mol$^{-1}$.
%
%
%The barriers among these conformational states are shown as P to Q, Q to P, Q to R and R to Q are calculated and compared. The convergence plots for these barriers for TASS and PBTASS shows that PBTASS(8ns) obtained convergence 2ns before TASS (10ns). The surface shows more symmetry when obtained from PBTASS method. This symmetry is expected because the system is symmetric. 

The one-dimensional projections of $F(\phi_1,\psi_1,\phi_2,\psi_2)$ along $\phi_1$ and $\phi_2$ were computed 
for the TASS and PBTASS cases, and are given in \fref{fig:tri-ala1.1}. 
$F(\phi_1)$ computed using TASS and PBTASS methods agrees well with each other, while $F(\phi_2)$ shows 
a difference of up to 3~kcal~mol$^{-1}$.
This indicates that some parts of the
high dimensional free energy landscape are not converged in TASS, as a result of the high $\Delta T$. 
%{\color{red} {\AAcomment need more clarity, I guess}}

%As mentioned earlier, this could be due to the high value of the 
%$\Delta T$ parameter in TASS. 
%was kept to a very high value to make a fair comparison with PBTASS results.
%
%Based on these data, we conclude that PBTASS performs better than the standard TASS method.

For further analysis, we projected the four-dimensional free energy surfaces to $(\phi_1,\psi_1)$, and $(\phi_2,\psi_2)$
spaces; See \fref{fig:tri-ala3} and {Figure~S1}. 
% Move phi_2,psi2 figure to SI.
%
%We compared free energy barriers, and relative free energies of saddle points.
%
Our main interest was not only to compare the convergence of barriers, but also to check the convergence of free energies of saddle points where the sampling is apparently poor.
We notice that the free energy barrier for {\bf C}$\rightarrow${\bf B} converges quickly and the results from PBTASS and TASS simulations are in good agreement.
The free energies of both {\bf T1} and {\bf T2} saddle points are nearly identical in PBTASS, while they deviate about 2~kcal~mol$^{-1}$ in TASS (\tref{table:ala3}).
%
%It is likely that this discrepancy is likely to arise due to difference in the extent of
%sampling in both methods, with PBTASS being more exhaustive than TASS.
%
The free energy of {\bf T3} is also nearly the same in both methods.
The highest energy saddle point {\bf T4} 
was found to converge quickly in PBTASS compared to TASS.
Similar observation can be also made while analyzing the free energy landscape
$F(\phi_2,\psi_2)$; See {Figure~S1}.
These results are assuring the accuracy and efficiency of PBTASS %over TASS 
in exploring high dimensional free energy landscapes.
%

%Ramachandran plots are also given in the figure ~\ref{fig:tri-ala3}. 
%
%The convergence along the CVs was checked for TASS and PBTASS method which is shown the figures \ref{fig:tri-ala1.0}, \ref{fig:tri-ala3} \ref{fig:tri-ala4}. 
%
%These pictures shows that PBTASS method samples high free energy surface much faster and better than TASS method.

\begin{figure*}
\centering
    \includegraphics[width=\textwidth]{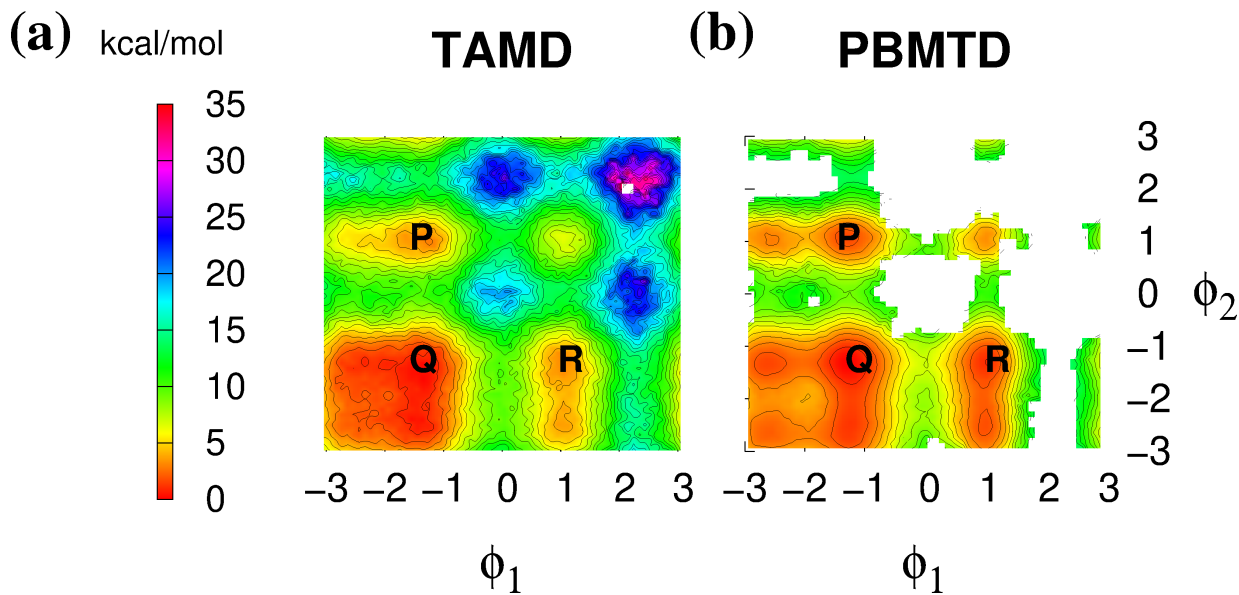}
    \caption{The free energy surface $F(\phi_1,\phi_2)$ of alanine tripeptide {\em in vacuo} computed from (a) TAMD and (b) PBMTD simulations after 20~ns are presented. Here angles are in radians. Contours are drawn for every 1 kcal~mol$^{-1}$.
    }
     \label{fig:tri-ala5}
\end{figure*}

Finally, we compare the performance of PBTASS with TAMD and PBMTD methods.
The free energy surface $F(\phi_1,\phi_2)$ obtained from the TAMD and PBMTD calculations
are shown in \fref{fig:tri-ala5}.
It can be clearly seen that, even after 20~ns, the exploration of the free energy surface is not as exhaustive as that observed in PBTASS and TASS simulations.
%
%results are also compared with standard temperature accelerated molcelar dynamics (TAMD) and standard parallel bias metadynamics (PB-MTD) methods to shows why PBTASS method is more useful than these methods. 
%
The TAMD and PBMTD free energy surfaces are more noisy and the free energy barriers \rxn{Q}{P} and \rxn{Q}{R}
are 10.0 and 9.0~kcal~mol$^{-1}$, respectively, with TAMD, while they are 9.8 and 7.3~kcal~mol$^{-1}$, respectively, with PBMTD (\tref{table:ala3}).
%
%FES along $(\phi_1,\phi_2)$ is shown sampled by TAMD and PB-MTD method. The PB-MTD method alone could not sample the high free energy surface. T
Although the exploration of high energy regions is
much better with TAMD than PBMTD, 
detailed analysis show that
the high energy regions of the landscape 
are not properly converged within 20~ns, unlike 
we observed in PBTASS runs. 
This is apparent in Figure~S2.
%SI~Figure 2 of Si is $F(\phi_1,\psi_1)$ of TAMD and PBMTD 
%
%
Clearly, in PBTASS runs, the system was able to diffuse through the entire four-dimensional CV-space in a more exhaustive manner compared to other methods. 
This capability of PBTASS can be ascribed to 
the inherent divide-and-conquer procedure invoked by the 
US bias, in addition to the 
boosted sampling rendered through the combination of 
high temperature and parallel bias.

% \nncomment{Make a table with 
 %free energy barriers for
% \rxn{Q}{P}, \rxn{Q}{R}, \rxn{P}{Q}
% \rxn{R}{Q}, free energies of T1, T2, T3, T4 for various methods: a) PBTASS, TASS, TAMD, PBMTD. 
 %In the caption, mention also that
% the data computed from the
% free energy surface obtained
% after 20ns of simulation.
% } 
\begin{table}[t]
    \centering
\begin{tabular}{ |c|c|c|c|c|c|c|c|c| }
\hline \hline 
Method & \multicolumn{4}{c |}{$\Delta F^\ddagger$}               & \multicolumn{4}{c|}{$\Delta F$}  \\ \cline{2-9} 
       & \rxn{Q}{P} & \rxn{Q}{R} & \rxn{P}{Q}   &  \rxn{R}{Q} & {\bf T1} & {\bf T2} & {\bf T3} & {\bf T4} \\ \hline
PBTASS & 9.5 & 9.5  & 6.5 & 6.5  &  9.4 & 10.0 & 18.8 & 23.1\\ \hline
TASS   & 8.5 & 9.0  & 6.0 & 6.0  & 9.3 & 11.6 & 19.3 & 25.3 \\ \hline
TAMD   & 10.0 & 9.0 & 7.0 & 5.5  & 9.0 & 7.0 & 18.0 & -  \\ \hline
PBMTD  &    9.8 &  7.3    &  7.3    & 6.0     & 6.2 & - & - & - \\ \hline\hline
\end{tabular}
\caption{Free energies barriers ($\Delta F^\ddagger$) and 
free energies ($\Delta F$) of the saddle points (compared to the free energy of {\bf A})
computed after 20~ns using PBTASS, TASS, TAMD, and PBMTD simulations.
Free energies are in kcal~mol$^{-1}$.
Dash symbol ($-$) indicates that the 
corresponding free energies could not be computed due to noise arising from  
poor sampling.
}
\label{table:ala3}
\end{table}

Clearly, these results show that 
PBTASS is as accurate and efficient as TASS, while outperforms TAMD, and PBMTD methods.

\subsection{Alanine Pentapeptide {\em In Vacuo}}
\begin{figure}[htbp]
\centering
    \includegraphics[width=0.7\textwidth]{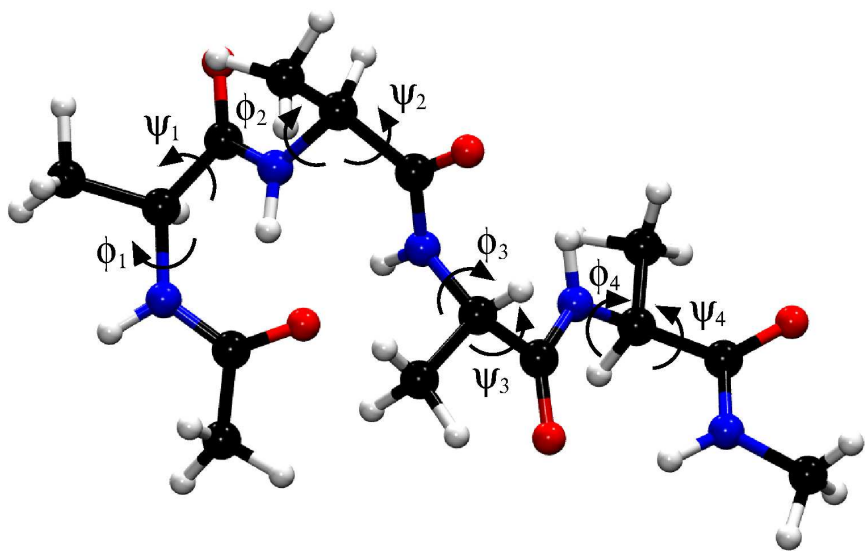}
     \caption{\label{penta-ala}
Alanine pentapeptide molecule is shown, where the eight Ramachandran angles that are taken as CVs are labeled.  }
     \end{figure}

As an application of PBTASS, we  carried out a detailed study of alanine pentapeptide {\em in vacuo}
aimed to compute the eight-dimensional free energy surface as a function of 
eight Ramachandran angles ($\phi_1,\psi_1,\phi_2,\psi_2, \phi_3, \psi_3, \phi_4, \psi_4$);
see \fref{penta-ala}.
MD calculations were performed using AMBER14\cite{ff14SB} %TODO Write the version number instead of XX
interfaced with PLUMED-2.2.3 \cite{plumed2.2.3}. 
We chose the ff14SB force field~\cite{ff14SB} for these simulations. 
%
%The free energy surface along eight torsional angles ($\phi_1,\psi_1,\phi_2,\psi_2, \phi_3, \psi_3, \phi_4, \psi_4$) as CVs  was sampled. Please look at the figure ~\ref{penta-ala} for CVs selection.

All the eight Ramachandran angles were taken as CVs in our PBTASS simulations.
Here we have arbitrarily opted the $\phi_2$ coordinate for applying the US bias.
The parallel (three-dimensional) MTD biases were applied along the other three $\phi$ angles, i.e. ($\phi_1, \phi_3, \phi_4$).
All the eight auxiliary variables were thermostatted at 3000~K. 
An over-damped Langevin thermostat with a friction coefficient of
0.1~fs$^{-1}$ was used to maintain their temperature.
%
%The performance of force field at this temperature was also checked by the deviation of their fluctuations of bonds, angles and dihedrals. 
The PBMTD bias potential was updated every 500~fs and
the PBMTD parameters $w_1(0) = w_2(0) = w_3(0) = 0.6$~kcal~mol$^{-1}$ and $\delta s =0.05$ radians were taken. 
The parameter $\Delta T$ was set to 45000~K. 
The umbrella potential was applied from $-\pi$ to $\pi$ at an interval of 0.2 radians with $\kappa_h = 1.2\times10^2 $ kcal~ mol$^{-1}$rad$^{-2}$,  $k_i=1.2\times 10^3$~kcal~mol$^{-1}$rad$^{-2}$, and  $\mu_i=50$~amu {\AA}$^2$ rad$^{-2}$,
for $h=1,\cdots,33$ and $i=1,\cdots,8$.
We used the same setups for doing
TASS, TAMD, and PBMTD simulations.
In TASS, we applied the US bias along $\phi_2$ and one-dimensional MTD
bias along $\phi_1$.
In PBMTD simulations, we chose the four $\phi$ angles as CVs.
%
%Here, {\tt \color{Red} $M=33$ }
%and {\tt \color{red} $n=8$}.

%\nncomment{In the following make changes similar to that in tripeptide case}
%
%{\color{red} All the auxiliary variables were thermostatted to 3000K. 
%
%Parallel biases were updated every 500 fs and the initial parameters $w_1, w_2, w_3 =$ 0.6 kcal/mol were taken to be same, $\delta s_1, \delta s_2, \delta s_3 =$ 0.05 radians were also taken to be same and the bias factor in well- tempered parameter was taken to be 16 for simulation. Umbrella potential was applied from $-\pi$ to $\pi$ to an interval of 0.2 radians with $\kappa_h = 1.2\times10^2 $ kcal~ mol$^{-1}$rad$^{-2}$,  $\kappa_\alpha=1.2\times 10^3$~kcal~mol$^{-1}$rad$^{-2}$, and the mass of auxiliary variables was taken as  $\mu_\alpha=50$~amu {\AA}$^2$ rad$^{-2}$. } 

\begin{figure*}
\centering
    \includegraphics[width=\textwidth]{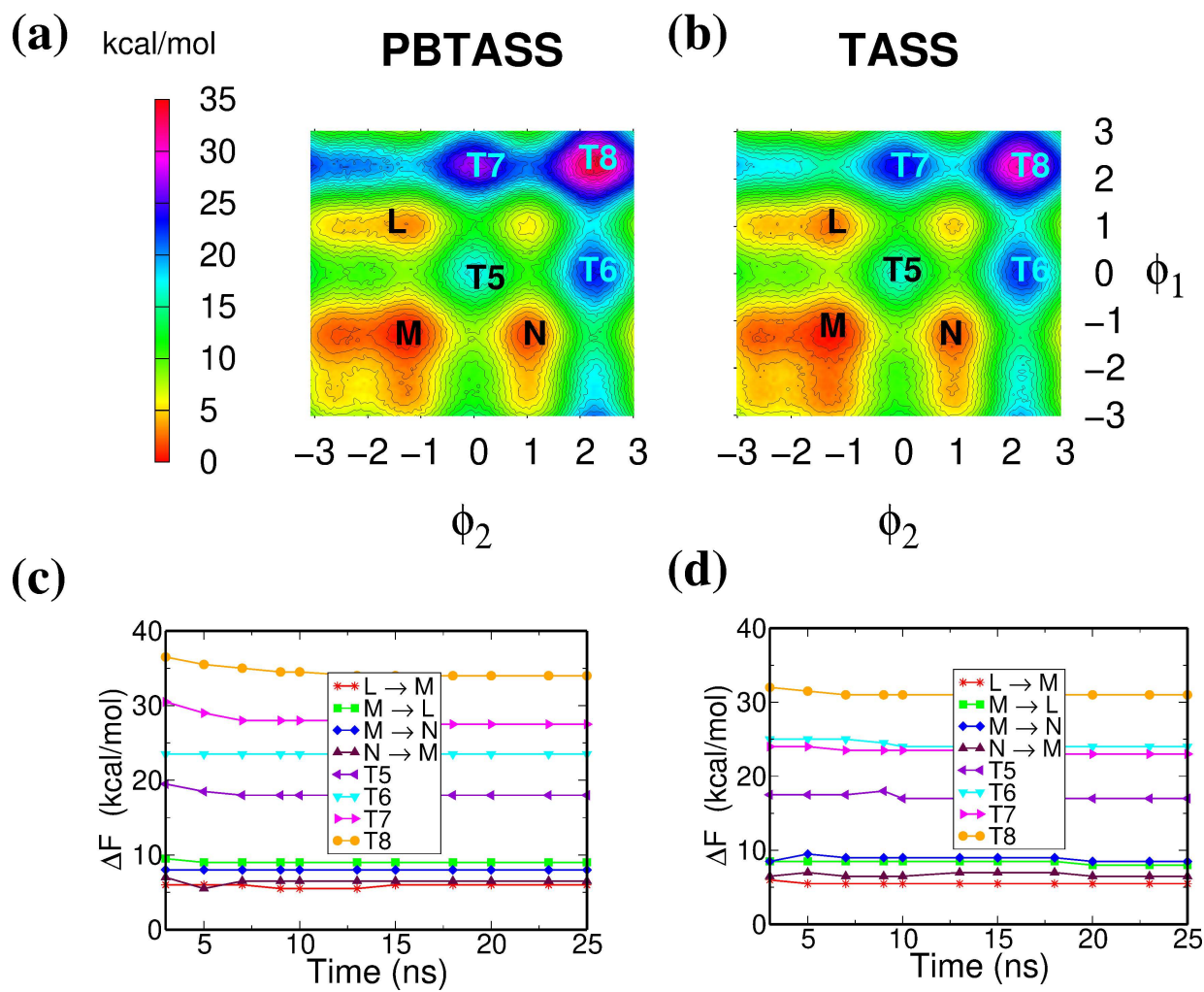}
    \caption{
    The free energy surface $F(\phi_1,\phi_2)$ 
    of alanine pentapeptide {\em in vacuo} computed
    from (a) PBTASS and (b) TASS simulations after 
    25~ns per window. 
    Some of the conformational states are labelled as {\bf L}, {\bf M}, {\bf N}, {\bf T5}, {\bf T6}, {\bf T7} and {\bf T8}. Contours are drawn for every 1~kcal~mol$^{-1}$. Convergence plots of some of the free energy barriers and free energies of saddle points with respect to the minimum {\bf M} as a function of  time per window are shown in the lower panels (c) and (d).  
    }
    \label{fig:penta-ala2}
\end{figure*}

\begin{figure*}
\centering
    \includegraphics[width=\textwidth]{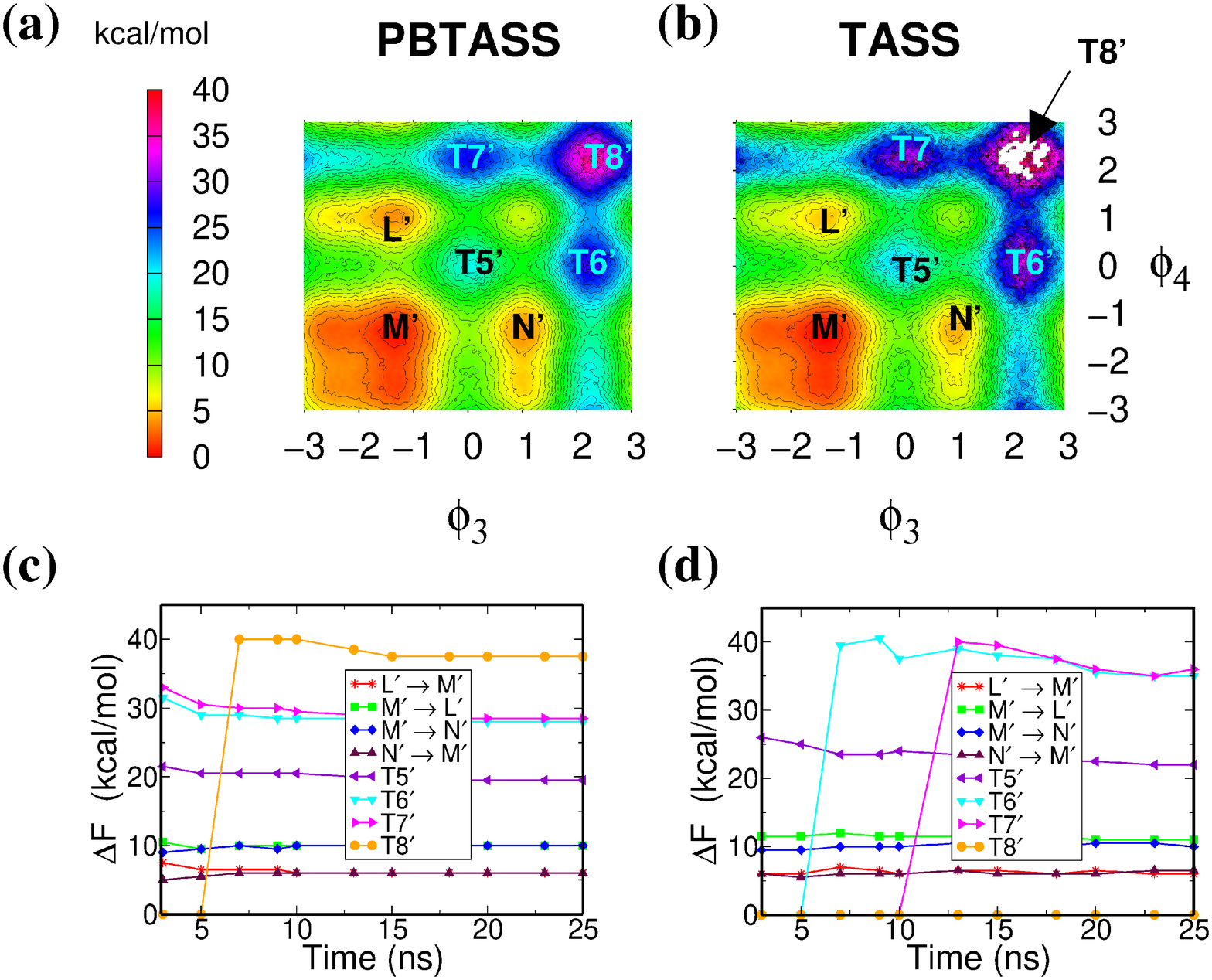}
    \caption{ 
    The free energy surface $F(\phi_3,\phi_4)$ 
    of alanine pentapeptide {\em in vacuo} computed
    from (a) PBTASS and (b) TASS simulations after 
    25~ns per window. 
    Some of the conformational states are labelled as {\bf L$^\prime$}, {\bf M$^\prime$}, {\bf N$^\prime$}, {\bf T5$^\prime$}, {\bf T6$^\prime$}, {\bf T7$^\prime$} and {\bf T8$^\prime$}. Contours are drawn for every 1~kcal~mol$^{-1}$. Convergence plots of some of the free energy barriers and free energies of saddle points with respect to the minimum {\bf M$^\prime$} as a function of  time per window are shown in the lower panels (c) and (d). Undetermined 
    free energies (due to poor sampling) are indicated as zero.
    }
    \label{fig:penta-ala6}
\end{figure*}

\begin{figure*}
\centering
    \includegraphics[width=\textwidth]{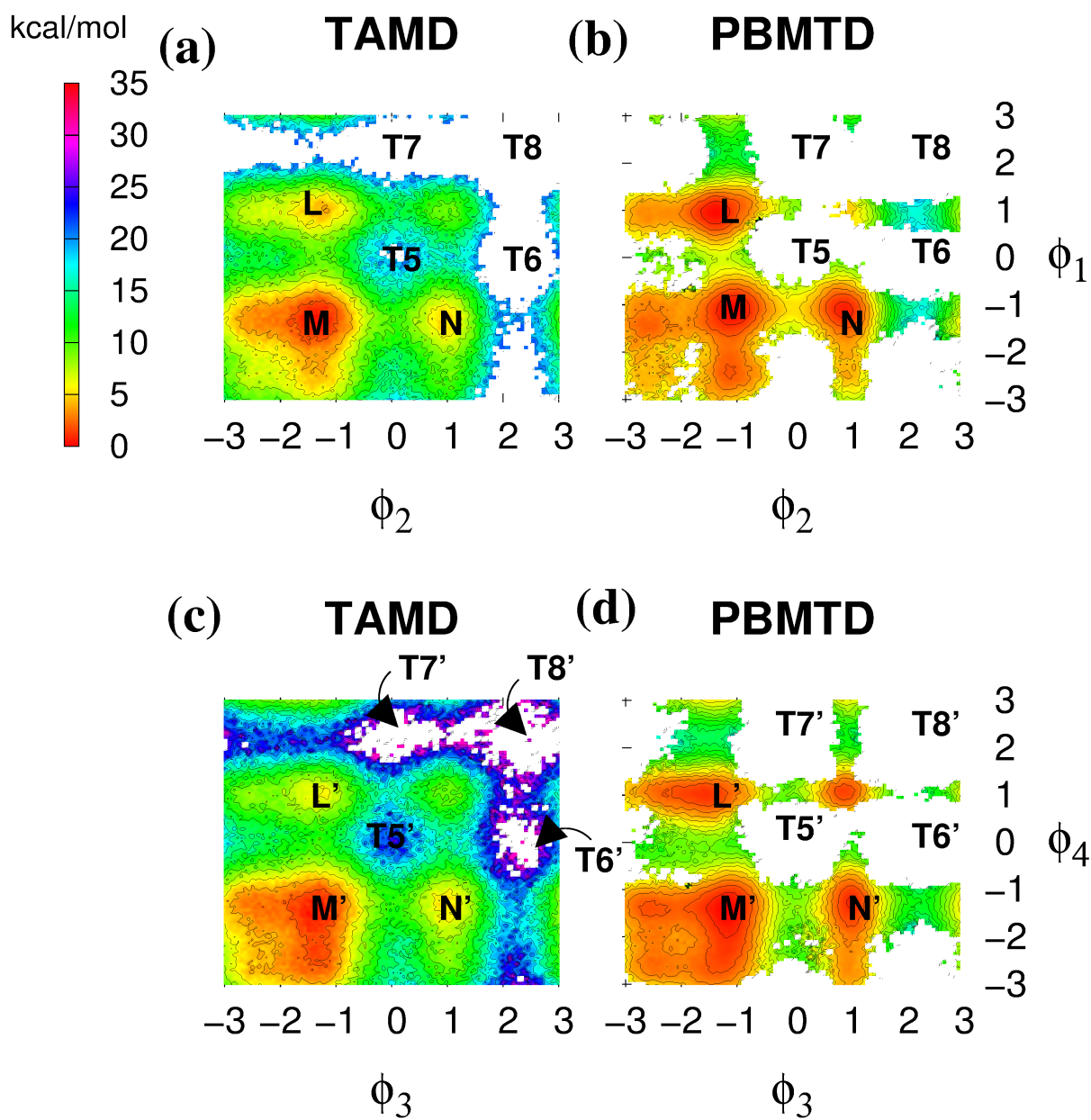}
    \caption{
    The free energy surfaces $F(\phi_1,\phi_2)$ (top panel) and $F(\phi_3,\phi_4)$ (bottom panel) of alanine pentapeptide {\em in vacuo} computed from TAMD (left panel) and PBMTD (right panel) simulations after 25~ns. Here angles are in radians. Contours are drawn for every 1 kcal~mol$^{-1}$. White patches on the surfaces show that the sampling 
    of these regions was not adequate enough for estimating free energies.
 %   \nncomment{$F(\phi_3,\phi_4)$ surfaces are missing! Labels are missing!}
    }
    \label{fig:penta-ala7}
\end{figure*}

%SI_Figure 3
%\begin{figure*}
%\centering
%    \includegraphics[width=\textwidth]{penta_ala_tamd_pbmtd_phi3_phi4_com.eps}
%    \caption{Alanine pentapeptide: FES along $(\phi_3,\phi_4)$ sampled using  (a) standard temperature accelerated molecular dynamics (TAMD) method and (b) standard parallel bias metadynamics (PBMTD) method after 25ns of simulation.  Every contour is plotted after 1 kcal/mol for PBMTD method and after 2 kcal/mol for TAMD method. The FES clearly shows that conformational states are not sampled well and are not differentiable.
%    \nncomment{make changes similar to the case of tripeptide}
%    }
%    \label{fig:penta-ala7}
%\end{figure*}

%SI_Figure 4
%\begin{figure*}
%\centering
%    \includegraphics[width=\textwidth]{1d_proj.eps}
%    \caption{Alanine pentapeptide: Comparison of 1D projected Free energy from 9D free energy obtained after 25ns of simulation along $\phi_1,\phi_2, \phi_3, \phi_4 $ sampled using standrad TASS and PBTASS methods.   }
%    \label{fig:penta-ala8}
%\end{figure*}

%
The high dimensional free energy surface obtained after 25~ns  (per window) of PBTASS simulation
was projected to ($\phi_1$,$\phi_2$) space and the convergence of the free energy barriers
were analyzed; See \fref{fig:penta-ala2}.
Considering the dimensionality of the explored surface, the extent of sampling and the observed smoothness of the projected surfaces are remarkable.
All the regions in the $(\phi_1,\phi_2)$ space, including all the 2$^{\rm nd}$ order saddles 
were sampled well.
Free energy barriers separating various 
metastable states
and the free energy of saddle points 
on this surface were computed and their convergence was analyzed.
%
%The computed free energy barriers for \rxn{Q}{R} and \rxn{Q}{P} are nearly the same and they converged quickly in PBTASS. 
%
The computed free energy barriers for \rxn{M}{L} and \rxn{M}{N} are nearly the same and they converged quickly in PBTASS.
The same was also seen for the reverse reactions; see also ~\tref{table:ala5}.
%
%\nncomment{include a table with free energy barriers on various surfaces}
%
A quick convergence of the free energies of the high energy saddle points was also observed
in the PBTASS simulation (\fref{fig:penta-ala2}).

The same analysis was also extended for TASS simulation.
Interestingly, free energy barriers computed from TASS runs are close to that
computed from PBTASS simulations (Table~\ref{table:ala5}, \fref{fig:penta-ala2}).
However, the free energies of some of the saddle points show large deviations from the PBTASS results.
%
%Figure XX shows the results of TAMD and TAMD simulations.\nncomment{figures should be in
%this order. First TAMD, and then TAMD}
%
%\mycomment{need to write after analyzing the figure. Free energy barriers to be
%computed as well.}

As next, the free energy surfaces $F(\phi_3,\phi_4)$ computed using PBTASS and TASS methods are compared (\fref{fig:penta-ala6}).
%
%Similar to that we observed in the case of $F(\phi_1,\phi_2)$, 
The convergence of 
free energy barriers 
%separating the major conformational states on these landscape 
are quick in both 
these
%PBTASS and TASS 
methods, and they agree well with each other.
%\mycomment{name P,Q,R differently in different surfaces}.
%
However, most striking difference between PBTASS and TASS can be seen in the convergence of the free energies of the saddle points.
Especially, the {\bf T8$^\prime$} saddle point was not sampled well enough to resolve in TASS, 
but the sampling was acceptable in PBTASS.
This lead us to conclude that the differences in the estimates of 
free energies of saddle points from PBTASS and TASS simulations are 
resulting from the poor exploration of saddle point regions in TASS compared to
PBTASS. 
Analyses of other projections of the free energy surface also yield the same conclusions
(See Figure~S3, S4, and  Table~S1). 
%
%TODO: SI~Figure~5 (a): F(phi_1,psi_1)  (b) F(phi_2,psi_2) (c) F(phi_3,psi_3), (d) F(phi_4,psi_4)
%                        computed from PBTASS only. No need to show convergence plot. The free energy surface at 25ns is to be only shown.
%
%
%TODO: SI~Table~1
% Free energy barriers for various minima on the above four free energy landscapes at 25 ns (only PBTASS results)

Free energy surfaces $F(\phi_1,\phi_2)$ and $F(\phi_3,\phi_4)$ computed using TAMD, and  PBMTD methods are shown in \fref{fig:penta-ala7}.
Both TAMD and PBMTD simulations were of 25~ns long.
It can be seen that low
free energy regions of
the CV-space were sampled 
reasonably well by both the methods.
However, some of the free energy barriers 
computed from TAMD and PBMTD
deviated up to 2~kcal~mol$^{-1}$
from the PBTASS estimates; See 
\tref{table:ala5}.
Free energies of several saddle points on the surfaces were not computable from the TAMD and PBMTD simulations as a consequence of
poor sampling.

These results show that the PBTASS approach outperforms the other three methods. 
The PBTASS method is able to thoroughly sample the eight-dimensional surface of alanine pentapeptide and could provide reliable free energy estimates within an affordable simulation time.
%We compared our results with the original TASS method, conventional temperature accelerated molecular dynamics (TAMD) and conventional parallel bias metadynamics (PB-MTD) method. The convergence was also checked for TASS and PBTASS method. \\
%More detailed analyses of these results are available in SI~Section~XX. 

%\nncomment{Make a table with 
% free energy barriers for
% \rxn{Q}{P}, \rxn{Q}{R}, \rxn{P}{Q}
% \rxn{R}{Q}, free energies of T1, T2, T3, T4 for various methods: a) PBTASS, TASS, TAMD, PBMTD. 
% In the caption, mention also that
% the data computed from the
% free energy surface obtained
% after 20ns of simulation.
% } 
% \AAcomment{The following table is for $F(\phi_2, \phi_1)$ surface}
 
 \begin{table}[t]
    \centering
\begin{tabular}{ |c|c|c|c|c|c|c|c|c| }
\hline \hline 
Method & \multicolumn{4}{c |}{$\Delta F^\ddagger$}               & \multicolumn{4}{c|}{$\Delta F$}  \\ \cline{2-9}
       & \rxn{M}{L} & \rxn{M}{N} & \rxn{L}{M} & \rxn{N}{M} & {\bf T5} & {\bf T6} & {\bf T7} & {\bf T8} \\ \hline \hline
PBTASS   & 8.9 & 8.0  & 5.9 & 6.6  & 17.8 & 23.5 & 27.5 & 34.0  \\ \hline 
TASS  & 8.0 & 8.5  &  5.5 & 6.9 & 16.7 & 22.9 & 24.1 & 31.1 \\ \hline 
TAMD & 9.4 & 10.3  &  5.9 & 5.9 & 23.6   & -    & -    & -     \\ \hline
PBMTD    & 5.2 & 6.6  & 4.6  & 6.5 & - & - & - & - \\ \hline \hline
%PBMTD   & 5.5 & 6.2  & 3.7  & 6.9 & - & - & - & - \\ \hline \hline 
         &  \rxn{M$^\prime$}{L$^\prime$} & \rxn{M$^\prime$}{N$^\prime$} & \rxn{L$^\prime$}{M$^\prime$}  &  \rxn{N$^\prime$}{M$^\prime$}  &  {\bf T5$^\prime$} & {\bf T6$^\prime$}  & {\bf T7$^\prime$} & {\bf T8$^\prime$} \\  \hline
PBTASS &  10.0 & 9.8 & 6.3 & 6.2 & 19.6 & 28.5 & 27.6 & 37.5  \\
\hline
TASS &  10.9 & 10.1 & 5.9 & 5.6 & 22.1 & 35.3 & 35.2 & - \\
\hline
TAMD &  11.5 & 11.0 & 6.1 & 5.6 & 25.0 & - & - &- \\
\hline
PBMTD  & 9.1 & 8.2 &  8.0 & 7.3 & - & - & - & - \\ \hline \hline
%PBMTD & 7.8 & 8.8 &  6.5 & 8.0 & - & - & - &- \\ \hline \hline 
\end{tabular}
\caption{Free energies barriers and free energies of the saddle points (compared to the free energy of {\bf M} or {\bf M$^\prime$}) from free energy surfaces  for alanine pentapeptide {\em in vacuo} as in Figures \ref{fig:penta-ala2} and \ref{fig:penta-ala6} computed using PBTASS, TASS, TAMD, and PBMTD methods after 25~ns. Free energies are in kcal~mol$^{-1}$.}
\label{table:ala5}
\end{table}

\section{Conclusions}

The PBTASS method introduced in this work 
combines the PBMTD high dimensional bias 
with the TASS Lagrangian. 
This brings a major boost in the  efficiency of TASS in sampling 
a high dimensional CV-space.
%
%In this method, we used the PBMTD bias potential together within the TASS Lagrangian. 
%
%The method permits one to use a high-dimensional MTD bias potential in TASS simulations.
%
We have demonstrated the accuracy and the efficiency of this method in exploring high dimensional 
free energy surfaces, and for free energy calculations by taking the examples of alanine tripeptide and alanine pentapeptide. 
%
%The high free energy saddle regions of the eight-dimensional free energy surface of alanine pentapeptide 
%were explored sufficiently well using PBTASS.
%

The advantage of the PBTASS method over TASS is that it can bias more number of CVs and can enhance the sweeping motion of the system in a high dimensional CV-space while retaining all the salient features of the original TASS method.\cite{Shalini:2017}. 
Thus, the PBTASS method is a promising alternative to
TASS, TAMD, PBMTD, and similar methods, 
for calculating free energies of chemical reactions and
structural transformations occurring in 
large soft matter systems.
%wherein more CVs need to be 
%biased.                                                   %for free energy calculations in 
%
%
Substantial progress has been made in using machine learning tools to 
determine order parameters in describing  rare-events.\cite{Tuckerman:ML:PRL:2017,Bonati:PNAS:2019,Noe:JCP:2018,Tiwary:JCP:2018,Ferguson:JCC:2018,Pande:JCP:2018}
Methods like PBTASS, which enables the system to transverse through a 
high dimensional CV-space in an exhaustive manner, are most suited to be
integrated with machine learning tools.\cite{Tuckerman:ML:2020}
This will be our focus in the near future.
%
%The variant provides us the flexibility to apply 1-dimensional MTD bias along any number of CVs unlike TASS where the MTD bias was limited to one or two CVs. Increasing the dimensionality of MTD bias limits the efficiency of TASS method but PBTASS has overcome this limitation of TASS. Now we can claim that there is a method that can enhance the sampling of any number of CVs.  

%\begin{acknowledgments}  % Commented by Shivani
\section*{Acknowledgments}  % edited by Shivani
Authors acknowledge the HPC facility (HPC2013) at the Indian Institute of Technology Kanpur.
%Computations were performed using the HPC cluster at Indian Institute of Technology Kanpur.
%
AG and SV thank the DST-INSPIRE for their Ph.D. fellowship.
%
%
%\end{acknowledgments}.  % Commented by Shivani

%\bibliography{references.bib}% Produces the bibliography via BibTeX.bibliography style of JCTC}
\bibliography{PBTASS.bib}
\end{document}